# Partial crystallization in a Zr-based bulk metallic glass in selective laser melting

R.S. Khmyrov,[a] P.A. Podrabinnik,[a] T.V. Tarasova,[a] M.A. Gridnev,[a] A.D. Korotkov,[a]
S.N. Grigoriev,[b] A.Yu. Kurmysheva,[b] O.B. Kovalev,[c] A.V. Gusarov[a*]

[a] *Laboratory of Innovative Additive Technologies, Moscow State University of Technology STANKIN, Vadkovsky Per. 3a, 127055 Moscow, Russia*
[b] *Laboratory of Electric Current Assisted Sintering Technologies, Moscow State University of Technology STANKIN, Vadkovsky Per. 3a, 127055 Moscow, Russia*
[c] *Khristianovich Institute of Theoretical and Applied Mechanics SB RAS, 630090 Novosibirsk, Russia*

[*] Corresponding author e-mail: av.goussarov@gmail.com

**Abstract.** Metals and alloys in amorphous state are promising for the use as structural and functional materials due to their superior properties related to the absence of such defects as grain boundaries and dislocations. Obtaining the amorphous state requires quenching from liquid state with a high cooling rate. In conventional technologies, this is the reason for a considerable size limitation hindering application of amorphous alloys. Additive manufacturing (AM) is free of the size limitation. The so-called bulk metallic glass (BMG) alloys have extremely low critical cooling rates and can be used in AM. Recent studies on AM from Zr-based BMGs by selective laser melting (SLM) has proved the possibility of attaining the amorphous state and revealed partial crystallization to be the principal drawback of this process. The present work aims to analyze the conditions for partial crystallization and try to control it by optimizing the process parameters. A comprehensive parametric analysis is accomplished in single-track SLM experiments with Zr-based BMG alloy Vit 106. The observed microstructures are related to the temperature fields and thermal cycles estimated by an analytic heat-transfer model in the laser-impact zone. In the cross section of a laser track, a central bright domain is identified as the remelted zone. An annular darker crystallization zone encircles the remelted zone. The model fits the experimentally obtained dependencies of the remelted zone size versus laser power and scanning speed and indicates that $43 \pm 2\%$ of the incident laser energy is transferred into the substrate thermal energy. It seems that primary crystalline inclusions existing in the substrate before laser processing dissolve at the laser melting. The mean cooling rate in the remelted zone is much greater than the critical cooling rate. Therefore, homogeneous nucleation is not expected. Nevertheless, the theoretically estimated crystallization times are sufficient for a considerable crystal growth in the heat-affected zone where primary crystalline inclusions and nuclei are not completely dissolved.

## 2. Introduction

Metal alloys in amorphous state have potentially higher mechanical properties, corrosion resistance, and some other functional properties compared to those alloys in crystalline state. The absence of such defects as grain boundaries and dislocations explains the superior properties. Obtaining an amorphous state generally requires quenching from liquid state with high cooling rates greater than the so-called critical cooling rate. The higher critical cooling rate the lower glass forming ability. Conventional structural materials like steels have low glass forming ability. The so-called bulk metallic glasses (BMG) are alloys commonly considered for manufacturing amorphous-state structures [1]. BMGs are alloys designed for high glass forming ability. Normally, BMG amorphous ingots can be obtained with the minimum of the three dimensions greater than one millimeter. In the current structural applications, Zr-based BMGs are the most promising among the known BMG alloys due to their excellent fracture toughness [2]. The examples of commercial application of such alloys are golf heads, guitar pins, and small mechanical parts for consumer electronics [2].

In conventional forming processes like casting, the cooling rate is controlled by heat transfer from the bulk of the part/ingot. Therefore, cooling rate generally decreases with the part size. Starting from a certain size, the maximum possible cooling rate attains the critical cooling rate, so that the given material cannot be obtained in amorphous state. The size limitation is very serious. To overcome it, new BMG alloys with higher glass forming ability are being developed. Sometimes, the alloys optimized for the minimum critical cooling rate do not have optimal functional properties. Another approach to increase the part size is the use of additive manufacturing where material is added to a growing part step-by-step. One can assure a high cooling rate of a small portion of the added material while the number of these portions is not limited. Thus, there is no size limit in additive manufacturing. Indeed, metal parts with amorphous structure have been successfully obtained by such well-known additive processes as selective laser melting (SLM), laser directed energy deposition, fused filament fabrication, and powder spraying [3].

It seems that selective laser melting (SLM) also known as laser powder-bed fusion (LPBF) has become the most commercialized additive technology applied for manufacturing from metals and alloys. The reason is the small spot < 100 μm of currently available technological lasers that assures a satisfactory precision, their high power sufficient to melt even refractory materials, and a large market of high-quality powders of metals and alloys [4]. A considerable experience has been accumulated on SLM from BMG alloys [5]. For example, such a massive article as a hip prosthesis was fabricated from a Zr-based BMG by SLM [6]. The cooling rate in SLM can be several orders of magnitude greater than the critical cooling rate of the processed BMG alloy. Surprisingly, partial crystallization of such alloys is often observed [7]. The common explanation is rather complicated non-monotonous dependence of temperature versus time due to multiple laser scans. For example, the first laser scan melts material in the given point and turns it into amorphous state while the second laser scan heats this portion of material to a temperature below the liquidus point and induces partial crystallization. Generally, the domains near laser-scanning lines are in amorphous state while domains covered by heat-affected zones can be in crystalline state. Thus, characteristic fish-scale amorphous/crystalline patterns are formed [7].

Controlling the crystallization in laser processing of BMG alloys can be advantageous for formation of a desirable amorphous/crystalline microstructure. However, in Zr-based alloys, crystallization is rather an undesirable process since even a small amount of crystals can significantly reduce the fracture toughness [8]. Understanding crystallization in such conditions requires further experimental [9] and theoretical [10] fundamental researches in thermodynamics and kinetics of BMG alloys. The fundamental researches promote development of predictive crystallization models [11,12] combining heat transfer and crystallization kinetics in SLM.

Conventional experimental studies of phase transformations and structural relaxation in amorphous materials concern simple conditions of monotonous heating or cooling [9]. The models of crystallization kinetics essentially use the categories of homogenous nucleation and crystal growth and assume quasi-isothermal conditions like the most known Johnson-Mehl-Avrami-Kolmogorov model [12]. Such models are often applied in the strongly unsteady conditions of laser processing characterized by important temperature gradients. The experimental study of BMG crystallization in SLM is often reduced to observation of the microstructure formed [7] while systematic experimental studies are limited. The present work aims to fill this gap.

The present work concerns single-track experiments is SLM. The microstructure is characterized for a wide variety of process parameters typical for SLM. Laser processes an amorphous substrate without addition of powder to simplify experiments and exclude additional parametes related to powder. However, such an approach cannot significantly influence the thermal cycles of laser processing, which are essential for microstructure formation. The obtained microstructures are related to the temperature fields and thermal cycles estimated by an analytic model of heat transfer in the laser-impact zone. Original experiments characterize the laser-processed material and provide the key thermophysical parameters for the model. Section 2 describes the experimental approaches. Section 3 describes the heat transfer model. Section 4 presents the experimental results. Section 5 analyses the experimental results with the use of the heat transfer model.

## 2. Material and experimental methods

Table 1 shows the reported by manufacturer PrometalTech [13] chemical composition tolerance for the studied Zr-based alloy. The composition is close to that of Vit106 alloy with nominal atomic composition formula $Zr_{57}Cu_{15.4}Ni_{12.6}Al_{10}Nb_5$ [14]. The atomic mass $m$ of the elements is given according to Ref. [15]. Table 2 reports literature data [16-18] on the following properties of this alloy: density $\rho$, glass transition $T_g$, onset crystallization $T_x$, solidus $T_s$, and liquidus $T_l$ temperatures. The onset and offset melting temperatures are obtained by differential thermal analysis [18]. These values correspond to the solidus and liquidus points [17], respectively. The density reasonably decreases with temperature.

**Table 1.** Chemical composition of the studied alloy, $c$ (at.%)

| Element | | Zr | Cu | Ni | Al | Nb | Reference |
|---|---|---|---|---|---|---|---|
| Atomic mass, $m$ (a.m.u.) | | 91.224 | 63.546 | 58.6934 | 26.9815 | 92.9064 | [15] |
| $c$ | Vit106 nominal | 57 | 15.4 | 12.6 | 10 | 5 | [14] |
| | Manufacturer's tolerance | 54.15÷59.71 | 14.63÷16.13 | 11.97÷13.20 | 9.5÷10.5 | 4.75÷5.24 | [13] |
| | EDS analysis | 59.0 ± 0.3 | 14.3 ± 0.2 | 11.9 ± 0.2 | 9.3 ± 0.1 | 5.5 ± 0.2 | Present work |

Above the room temperature, the specific heat of a similar Zr alloy approximately follows the Dulong-Petit law in both amorphous and crystalline states [19]. This is reasonable because the Debye temperature of Zr is lower than the room temperature [20] and means that the specific heat per atom is equal to three Boltzmann constants $k$. It is supposed that the Dulong-Petit law is applicable to Vit 106 alloy too. Then, the mass specific heat is

$$C_p = \frac{3k}{c_{Zr}m_{Zr} + c_{Cu}m_{Cu} + c_{Ni}m_{Ni} + c_{Al}m_{Al} + c_{Nb}m_{Nb}}, \qquad (1)$$

where $c$ is the atomic fraction and $m$ the atomic mass according to Table 1 and the subscripts designate chemical elements. Table 2 shows the value calculated by Eq. (1).

**Table 2.** Properties of Vit106 alloy

| Property | | Value | Reference |
|---|---|---|---|
| Density, $\rho$ (g/cm$^3$) | at the room temperature, amorphous structure | 6.81 | [2.4] |
| | at the liquidus temperature | 6.520 | [2.5] |
| Temperature (K) | glass transition, $T_g$ | 672 | [2.6] |
| | | 690.6 | Thermal expansion |
| | onset crystallization, $T_x$ | 738 | [2.6] |
| | | 712.4 | Thermal expansion |
| | | 712 | Thermal diffusivity |
| | onset melting | 1092 | [2.6] |
| | solidus, $T_s$ | 1092 | [2.5] |
| | offset melting | 1120 | [2.6] |
| | liquidus, $T_l$ | 1115 | [2.5] |
| Specific heat, $C_p$ (J/(kg K)) | | 325.961 | Calculated |
| Linear thermal expansion coefficient, $\delta$ (K$^{-1}$) | | 1.1·10$^{-5}$ | Thermal expansion |

The raw material is supplied as 135×40 mm$^2$ plates of 2 mm thick with amorphous structure and possible crystalline inclusions. The chemical composition is measured by energy dispersion analysis of characteristic X-ray spectra (EDS) in scanning electron microscope (SEM) VEGA 3 LMH. The phase composition is studied by light microscope Olympus BX51M and the abovementioned SEM and X-ray diffractometer Panalytical Empyrean. Thermal expansion of the raw material is studied by dilatometry with NETZSCH DIL 402 in the thermal cycle including heating from the room temperature at the rate of 3 K/min up to 943 K followed by cooling at the same rate. Thermal diffusivity is measured by a laser flash method with NETZSCH LFA 467 in the thermal cycle consisting of stepwise heating from the room temperature up to 923 K with temperature stabilization at the measurement points followed by stepwise cooling. The samples for dilatometry 2×2×50 mm and laser flash 10×10×1 mm are cut by electrical discharge machining. Section 4 reports the results on the chemical and phase composition and the thermal expansion and laser-flash tests of the raw material.

Phase transformations in metallic glass are studied after laser treatement in the conditions of selective laser melting (SLM). The so-called single-track experiments are applied where the sample is scanned by a laser beam over well-separated lines to avoid overlapping the heat-affected zones of the laser-treated bands (tracks). Figure 1 shows the principal scheme of experimental SLM setup "ALAM" used for the singe-track experiments. A fiber laser generates near-infrared radiation. A galvanometer scanner focuses and moves a laser beam over the sample surface. The sample is inside a chamber with argon flow to avoid oxidation at elevated temperatures in the zone where the laser beam strikes the sample, the laser-impact zone. Table 3 summarizes laser processing parameters. They are typical in SLM. Samples for single-track experiments are 10×20×2 mm. The top surface is grinded, not polished to increase the contrast of laser-melted domains in the top view. In the SLM process, the laser beam scans a solid substrate covered with a thin powder layer, typically around 50 μm thick. This work does not apply powder because the thin layer cannot significantly change heat transfer in the laser-impact zone, and thus influence the thermal cycle of metallic glass determining phase transformations therein.

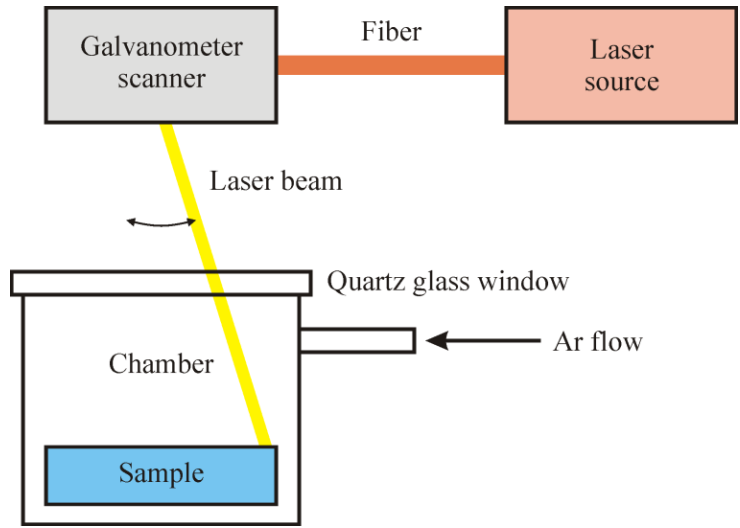

**Fig. 1.** Laser processing experiments.

**Table 3.** Parameters of single track experiments

| Parameter | Value |
|---|---|
| Laser wavelength | 1.07 μm |
| Laser spot diameter | 100 μm |
| Radiative laser energy | 20 ÷ 160 W |
| Scanning speed | 5 ÷ 3000 mm/s |

We observe the laser-treated samples from the top by light microscopy and in the cross sections perpendicular to the scanning direction by light microscopy and SEM. The cross sections are grinded and polished. To increase the contrast for light microscopy, some cross sections are etched with a mixture of hydrofluoric and nitric acids.

## 3. Model of heat transfer in the laser-impact zone

The structure of bulk metallic glass forms at temperatures below the liquidus temperature where the principal mode of heat transfer is conduction. The transient temperature field in the heat-affected zone around the laser beam can be estimated by the problem of a point source on the surface of a conductive half space [21]. If the laser beam scans the surface with a constant speed $V$, a steady-state temperature distribution $T(X,Y,Z)$ is formed [21] in the frame moving with the beam as shown in Fig. 2,

$$T - T_a = \frac{AP}{2\pi\lambda R} \exp\left(\frac{VX}{2\alpha} - \frac{VR}{2\alpha}\right), \qquad (2)$$

where $T_a$ is the ambient temperature, $P$ the laser beam power, $A$ the effective absorptivity, $\lambda$ the thermal conductivity, $\alpha$ the thermal diffusivity, and $R$ the distance from the intersection of the beam axis with the surface (see Fig. 2), $R^2 = X^2 + Y^2 + Z^2$. Effective absorptivity $A$ is the fraction of the laser beam radiative energy transferred into the heat energy of condensed phase and accounts for radiation reflection and other possible energy losses from the laser spot region as evaporation and thermal radiation [22]. Reference [22] demonstrated the applicability of the temperature field given by Eq. (2) in the conditions of selective laser melting by comparison with the observed microstructure.

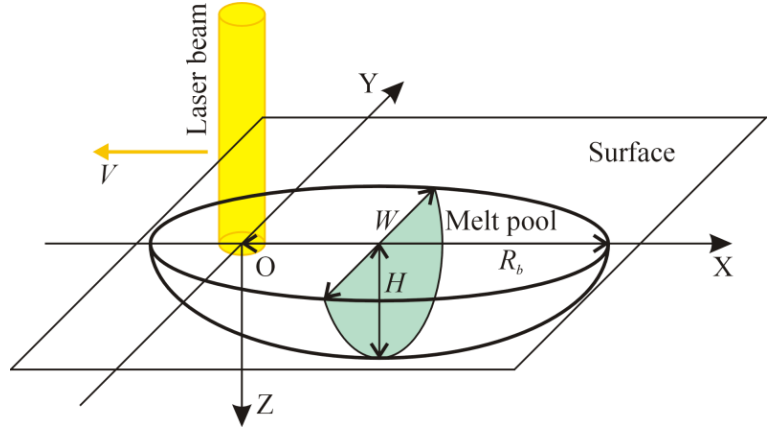

**Fig. 2.** Melt pool around the laser beam scanning with speed *V* in the direction opposite to axis (OX). Frame (OXYZ) moves with the beam. Parameters of the melt pool: depth *H*, width *W*, and distance $R_b$ from the laser spot center to the back side along axis (OX).

Along the scanning axis (OX) (see Fig. 2) behind the laser spot, $R = X$. Therefore, the exponential term vanishes in Eq. (2) and the temperature profile becomes independent of *V*. This is useful to define the characteristic distance

$$R_b = \frac{AP}{2\pi\lambda(T_m - T_a)}, \tag{3}$$

corresponding to a characteristic temperature $T_m$. This is the maximum distance from the point source to isotherm $T_m$ [23]. To simplify further analysis, Eq. (2) is written in the dimensionless form

$$\psi = \frac{1}{r}\exp(\Pi x - \Pi r), \tag{4}$$

with dimensionless temperature

$$\psi = \frac{T - T_a}{T_m - T_a}, \tag{5}$$

coordinates

$$(x, y, z, r) = \frac{(X, Y, Z, R)}{R_b}, \tag{6}$$

and thermal Peclet number $\Pi$ defined through characteristic velocity $V_0$,

$$\Pi = \frac{V}{V_0}, \quad V_0 = \frac{2\alpha}{R_b}. \tag{7}$$

It should be noted that according to Eq. (7), the product of the characteristic distance and characteristic velocity $R_b V_0 = 2\alpha$ is independent of laser processing parameters and is the property of the processed material.

At a constant value of $\psi$, Eq. (4) defines an isotherm, which is a surface of revolution around scanning axis (OX) (see Fig. 2). The generatrix profile $\rho(x)$, where $\rho^2 = y^2 + z^2$ is the dimensionless distance from axis (OX), is specified as a parametric curve

$$x = r + \frac{\ln(\psi r)}{\Pi}, \quad \rho = \sqrt{r^2 - x^2}, \tag{8}$$

with variable parameter $r$. The maximum distance of the isotherm from axis (OX), $\rho_m$, is found from the condition of local extremum, $d\rho/dx = 0$. The resulting dependence $\rho_m(\Pi)$ in parametric form is

$$\Pi = -\frac{\ln(\psi r)}{r[1 + \ln(\psi r)]}, \quad \rho_m = r\sqrt{1 - \ln^2(\psi r)}. \tag{9}$$

One can obtain from Eq. (9) that

$$\rho_m \to \frac{1}{\psi} \text{ at } \Pi \ll 1, \quad \rho_m \to \sqrt{\frac{2}{e\psi\Pi}} \text{ at } \Pi \gg 1, \tag{10}$$

where $e$ is Euler's number. The distance $\rho_m$ is the radius of the half-circle in the (XY) plane where the dimensionless temperature attains $\psi$, see the shadowed region in Fig. 2. Such a region with width $W$ and height $H$,

$$W = 2\rho_m R_b, \quad H = \rho_m R_b, \tag{11}$$

corresponds to a microstructure transformation at temperature $\psi$, which can be visible in the cross sections of laser-treated samples. Figure 3 shows the dimensionless curve $\rho_m(\Pi)$ for $\psi = 1$ estimating the depth and width of the remelted domain if $T_m$ is the liquidus temperature.

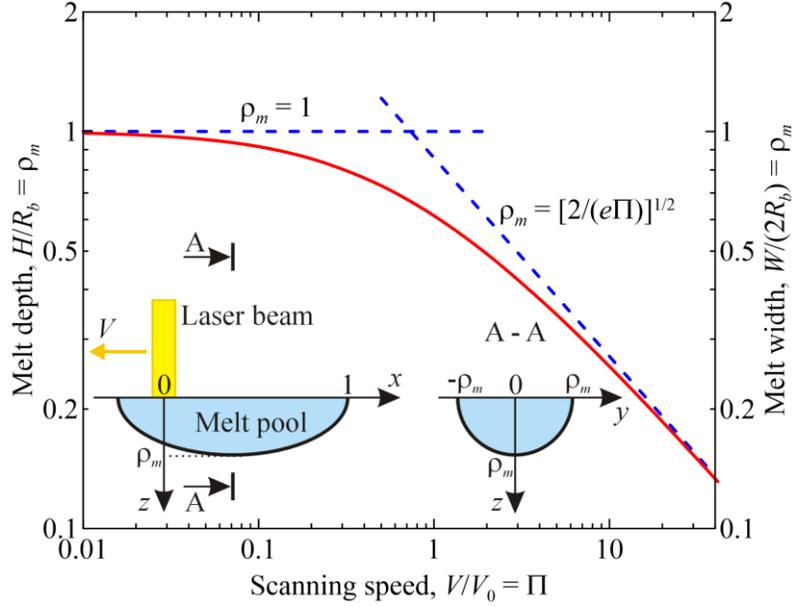

**Fig. 3.** Curve of dimensionless melt radius $\rho_m$ versus the thermal Peclet number $\Pi$ (solid red line) and its asymptotes (dashed blue lines). The insert shows the longitudinal section (left) and the maximum transversal section (right) of the melt pool in dimensionless coordinates $(x,y,z)$.

## 4. Results

### 4.1. Chemical and phase composition of raw material

The chemical composition of raw $Zr_{57}Cu_{15.4}Ni_{12.6}Al_{10}Nb_5$ alloy is measured by EDS analysis. Three cross sections are cut from different areas of a plate as shown in Fig. 4. In the bottom of Fig. 4, the insert shows three rectangular sampling domains (blue) where the EDS spectra are collected in every cross section. Two domains are near the plate surfaces while the third domain is in the middle. Thus, the composition is measured in nine sampling domains from different parts of the plate. We do not observe a variation of material composition with the position of the sampling domain inside the plate. Therefore, all the nine measurements are averaged. The last row of Table 1 shows the measured composition in the following format: "mean" ± "mean square deviation". The measured chemical composition essentially corresponds to the nominal Vit 106 composition and the manufacturer's tolerance listed in the same table.

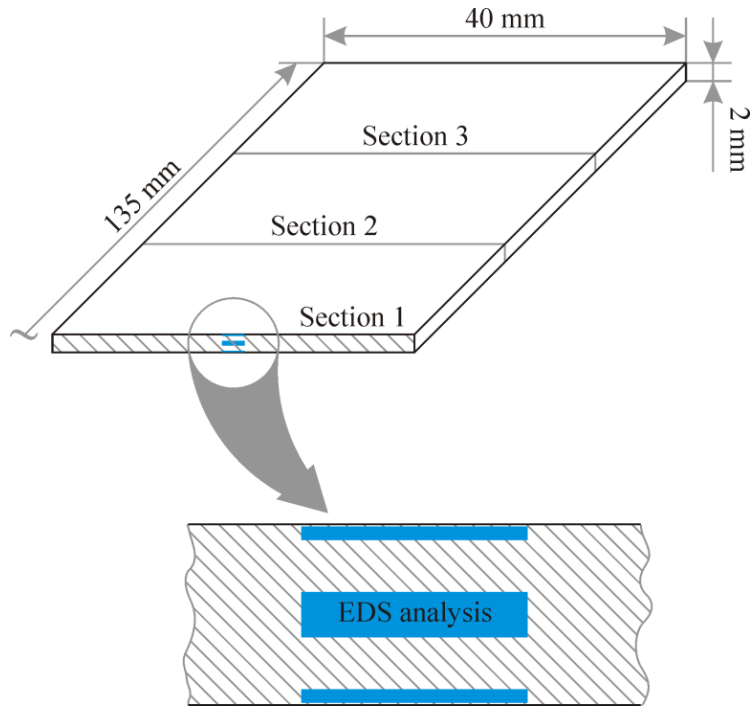

**Fig. 4.** Samples for EDS measurement of chemical composition. The blue rectangles indicate sampling domains in every cross section.

Figure 5 clearly distinguishes a matrix phase with few-micron inclusions in the microstructure observed both by light microscopy and SEM. The observed volume fraction of the inclusions is roughly 10%. Besides, SEM reveals inclusions of at least two contrast levels that can correspond to different phases. Generally, the darker phase contains heavier elements.

Figure 6 shows the X-ray diffraction (XRD) spectrum revealing a pronounced amorphous halo and a system of peaks from crystalline phases. According to Shadowspeaker *et al.* [24], there are five equilibrium crystalline phases in the considered system: $NiZr_2$, $CuZr_2$, $CuZr$, $Al_2Zr_3$, and $Al_3Zr_4$. Comparison with the reference spectra of the listed compounds [25] indicates that only $CuZr_2$ peaks match our spectrum, see vertical blue bars in Fig. 6, where the bar height is proportional to the intensity of the reference spectral line. The reference $CuZr_2$ peaks are slightly shifted to greater angles relative to the observed peaks. The reason can be a non-stoichiometry composition or admixture of other elements. The other four equilibrium phases are not revealed. The raw material is formed by quenching from liquid phase. Probably, there is no time sufficient for formation of these four phases. The chemical composition of these phases considerably deviates from the average composition. Thus, diffusion-controlled formation of them is likely inhibited. Several compounds with the composition around the average composition are checked too. Figure 6 shows that there are traces of $AlZr_3$ crystals while formation of a solid solution based on crystalline Zr is hardly possible. In summary, the probable crystalline inclusion phases are $CuZr_2$ and $AlZr_3$. The two compounds explain the inclusions of two contrast levels in Fig. 5b.

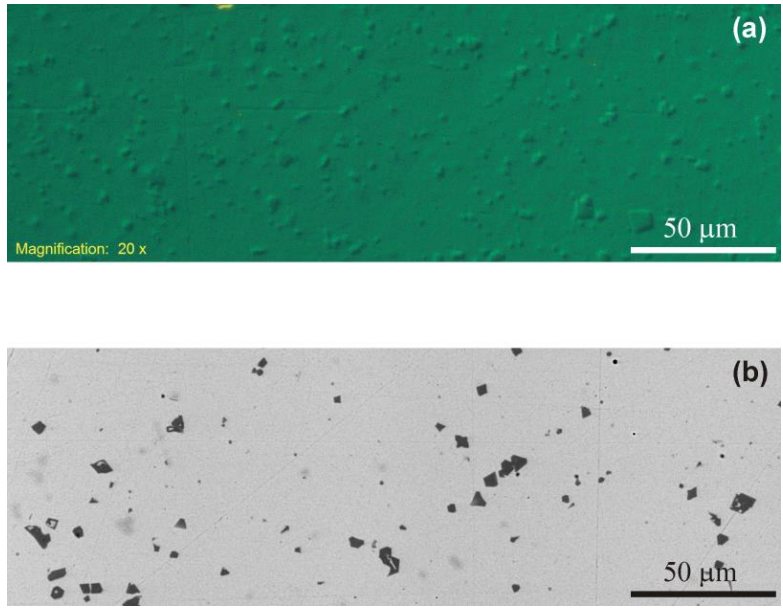

**Fig. 5.** Raw alloy microstructure: (a) light microscopy in dark field and (b) SEM in backscattered electrons.

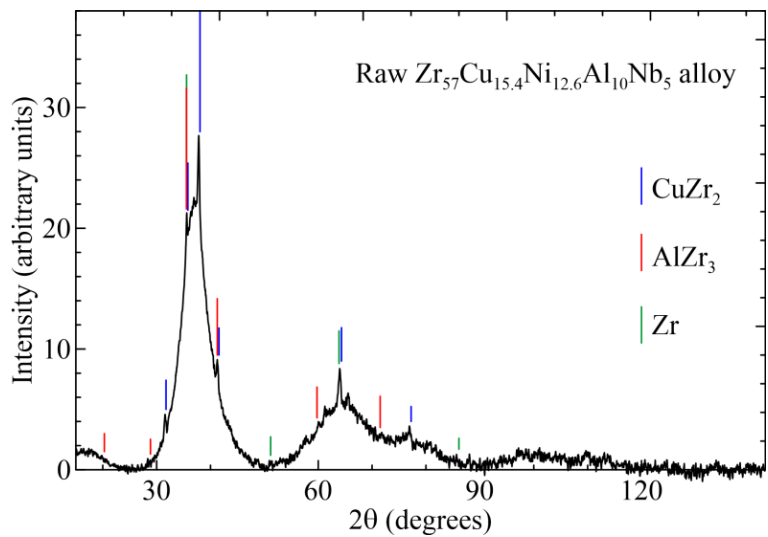

**Fig. 6.** Raw alloy Cu $K_\alpha$ XRD spectrum. The vertical bars point to the peaks of probable crystalline phases.

*4.2. Dilatometry of raw material*

Figure 7 shows linear thermal expansion of the raw material. Table 2 lists the glass transition and onset crystallization temperatures determined from the heating branch (red line) of the dilatometry curve. The glass transition point overestimates the literature data, see Table 2. This may reflect difference in chemical composition. On the contrary, the crystallization point underestimates the literature data. Crystallization may start earlier because of the crystalline inclusions. The inclusions can be nucleation centers, thus accelerating crystallization. The cooling branch (blue line) is approximately linear indicating no significant structure transformations. The linear thermal expansion coefficient derived from the cooling branch is approximately $\delta = 1.1 \cdot 10^{-5}$ $K^{-1}$. The slope of the heating branch below crystallization onset is approximately the same as the slope of the cooling branch. Thus, the thermal expansion coefficient is approximately the same in the amorphous and crystalline states.

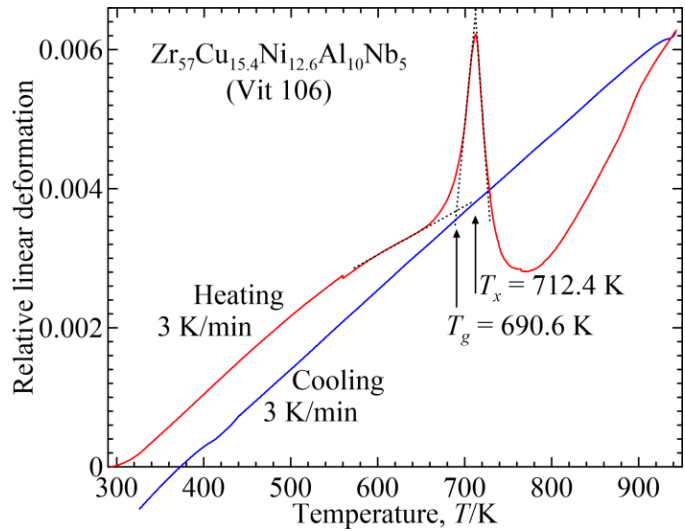

**Fig. 7.** Raw alloy dilatometry: heating (red line) and cooling (blue line).

*4.3. Thermal diffusivity of raw material*

Figure 8 shows the measured thermal diffusivity $\alpha$ of the raw material. The step on the heating branch (red) corresponds to the onset crystallization temperature $T_x$. The value of $T_x$ derived from the thermal diffusivity approximately corresponds to this value derived from the thermal expansion, see Table 2, but underestimates literature data on Vit 106 alloy listed in Table 2 too. The reason of the discrepancy with the literature data can be the crystalline inclusions in the studied raw material. These inclusions can promote heterogeneous crystallization. The cooling branch (blue) is significantly above the heating branch. It is supposed that the material essentially crystallizes during heating. Free electrons are principal heat carriers in metals. Metallic crystals are known to be more conductive than amorphous metals because free electrons are less scattered by regular atomic lattices. The cooling branch is more smooth than the heating one because there are no significant structure transformations at cooling.

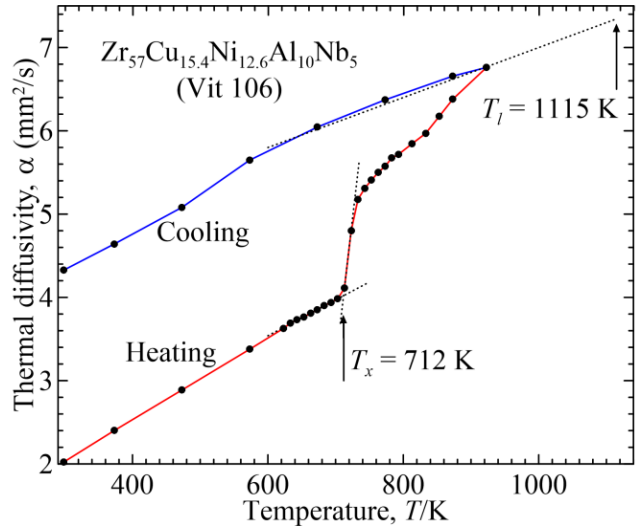

**Fig. 8.** Raw alloy thermal diffusivity: heating (red line) and cooling (blue line). The arrows point to onset crystallization $T_x$ and liquidus $T_l$ temperatures.

One can calculate thermal conductivity $\lambda$ from the thermal diffusivity as

$$\lambda = \rho C_p \alpha, \tag{12}$$

where the value of $C_p$ is taken from Table 2 and functions $\rho(T)$ for crystalline and amorphous states are estimated by extrapolation from the values of $\rho$ at the room $T_r = 298$ K and liquidus $T_l$ temperature, respectively, see Table 2. Table 4 shows the resulting functions. Figure 9 shows thermal conductivity $\lambda$ calculated by Eq. (12) and Table 4 from thermal diffusivity $\alpha$. The heating branch in Fig. 8 below $T_x$ results in the amorphous-state thermal conductivity in Fig. 9 while the cooling branch in Fig. 8 corresponds to the crystalline-state thermal conductivity in Fig 9. Thermal conductivity in the crystalline state is considerably greater than that in the amorphous state because electrons are less scattered by atoms arranged in a regular crystalline lattice. In Figs. 8 and 9, the dotted lines extrapolate the thermal diffusivity and conductivity to the onset crystallization and liquidus temperatures. Table 5 summarizes the resulting values at crystallization onset and melting. The indicated uncertainty is estimated based on observed measurement repeatability. The elevated uncertainty in the values of the last row is due to a considerable extrapolation interval.

**Table 4.** Estimated density $\rho$ versus temperature $T$

| Crystalline state | $\dfrac{\rho}{\text{g/cm}^3} = 6.81[1 - 3\delta(T - T_r)]$ |
|---|---|
| Amorphous state | $\dfrac{\rho}{\text{g/cm}^3} = 6.52[1 - 3\delta(T - T_l)]$ |

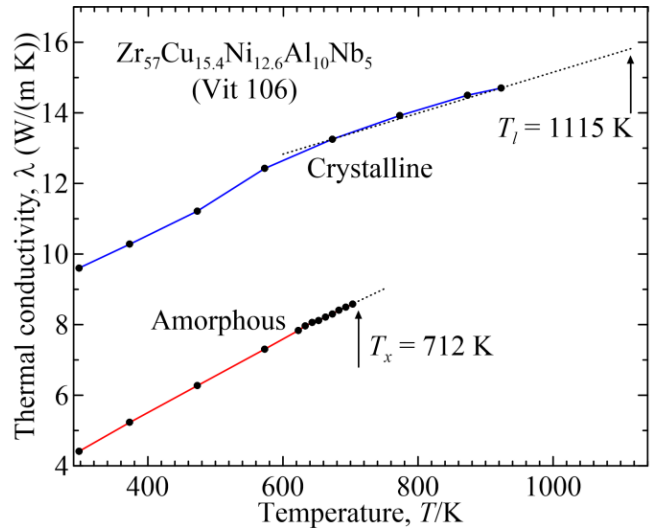

**Fig. 9.** Thermal conductivity of $Zr_{57}Cu_{15.4}Ni_{12.6}Al_{10}Nb_5$ alloy: amorphous state (red line) and crystalline state (blue line). The arrows point to onset crystallization $T_x$ and liquidus $T_l$ temperatures.

**Table 5.** Thermal diffusivity and conductivity of $Zr_{57}Cu_{15.4}Ni_{12.6}Al_{10}Nb_5$ alloy in amorphous and crystalline states at crystallization onset and melting

|  | Thermal diffusivity, $\alpha$ (mm²/s) | Thermal conductivity, $\lambda$ (W/(m K)) |
|---|---|---|
| Amorphous at crystallization onset | 4.0 ± 0.1 | 8.7 ± 0.1 |
| Crystalline at crystallization onset | 6.2 ± 0.1 | 13.5 ± 0.1 |
| Crystalline at melting | 7 ± 1 | 16 ± 2 |

*4.4. Microstructure after laser processing*

Figure 10 presents examples of microstructure formed by laser processing with the indicated parameters, laser beam power *P* and scanning speed *V*. The typical feature of the images is a brighter zone formed around the scanning axis, an imaginary line drawn by the laser beam axis on the sample surface. This zone looks as a band on the top views and a semicircle in the cross sections. Thus, this zone is a half cylinder with the axis being the scanning axis. This zone contains considerably less inclusions compared with the non-processes material. One can compare with the microstructures in Fig. 5 as well as non-processed domains in Fig. 10. Inclusions dissolve at melting. Therefore, the bright-contrast central zone can be identified as the remelted domain. This corresponds well to the theoretical prediction of the semicircle cross section of the melt pool shown in Fig. 2.

Bands of a darker contrast surround the remelted zone in Fig. 10. At the lower scanning speeds of 5 and 10 mm/s, these bands consist of columnar crystals observed on the top views. Generally, there are two rows of columnar crystals. In the inner one, the crystals are perpendicular to the scanning axis while they are oblique in the outer row. There can be a considerable space between the two rows of columnar crystals filled in with approximately equiaxed inclusions. See, for example, the top view at $V = 10$ mm/s in Fig. 10c. The columnar crystals generally grow in the direction of temperature gradient. Temperature decreases with the distance from the laser beam. Therefore, the temperature gradient vector approximately points to the center of the laser spot, the intersection of the laser beam with the surface. Thus, one can say that the columnar crystals roughly point to the laser spot center at the instant of their growth. This indicates that the inner columnar crystals form at the moment when the laser spot is at the minimum distance to them while the outer row of columnar crystals forms in a certain delay when the laser spot has moved forward.

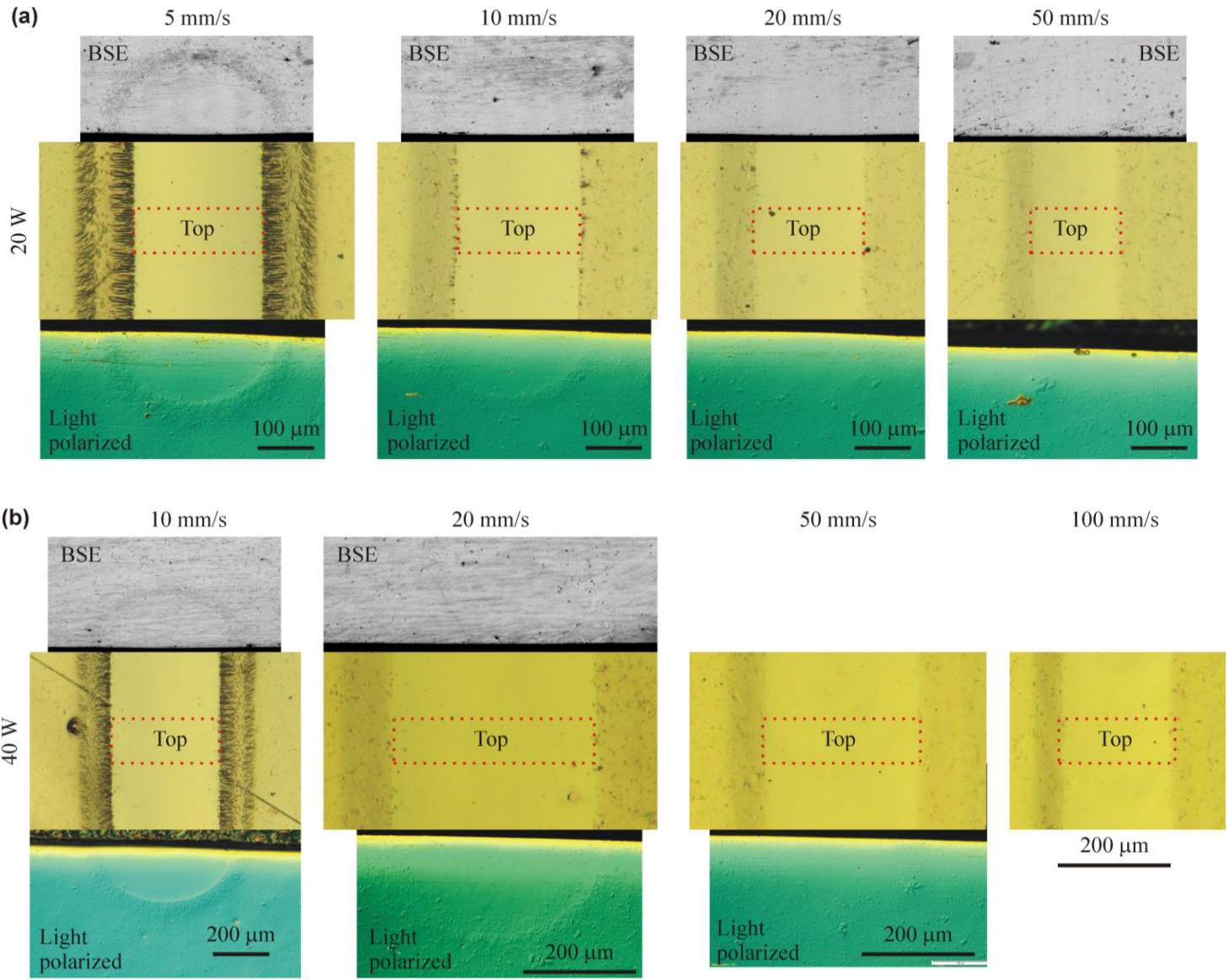

**Fig. 10.** Microstructure of laser-processed samples at laser power *P* (W): 20 (a), 40 (b), 80 (c), and 160 (d). Scanning speed *V* is indicated on the top. Visualization methods: top views in light microscope (Top) and cross sections perpendicular to the scanning direction obtained by: SEM in backscattered electrons (BSE), light microscopy in polarized light (Light polarized), and light microscopy after etching (Light etched). The dotted-line rectangles indicate the remelted width *W* on the top views.

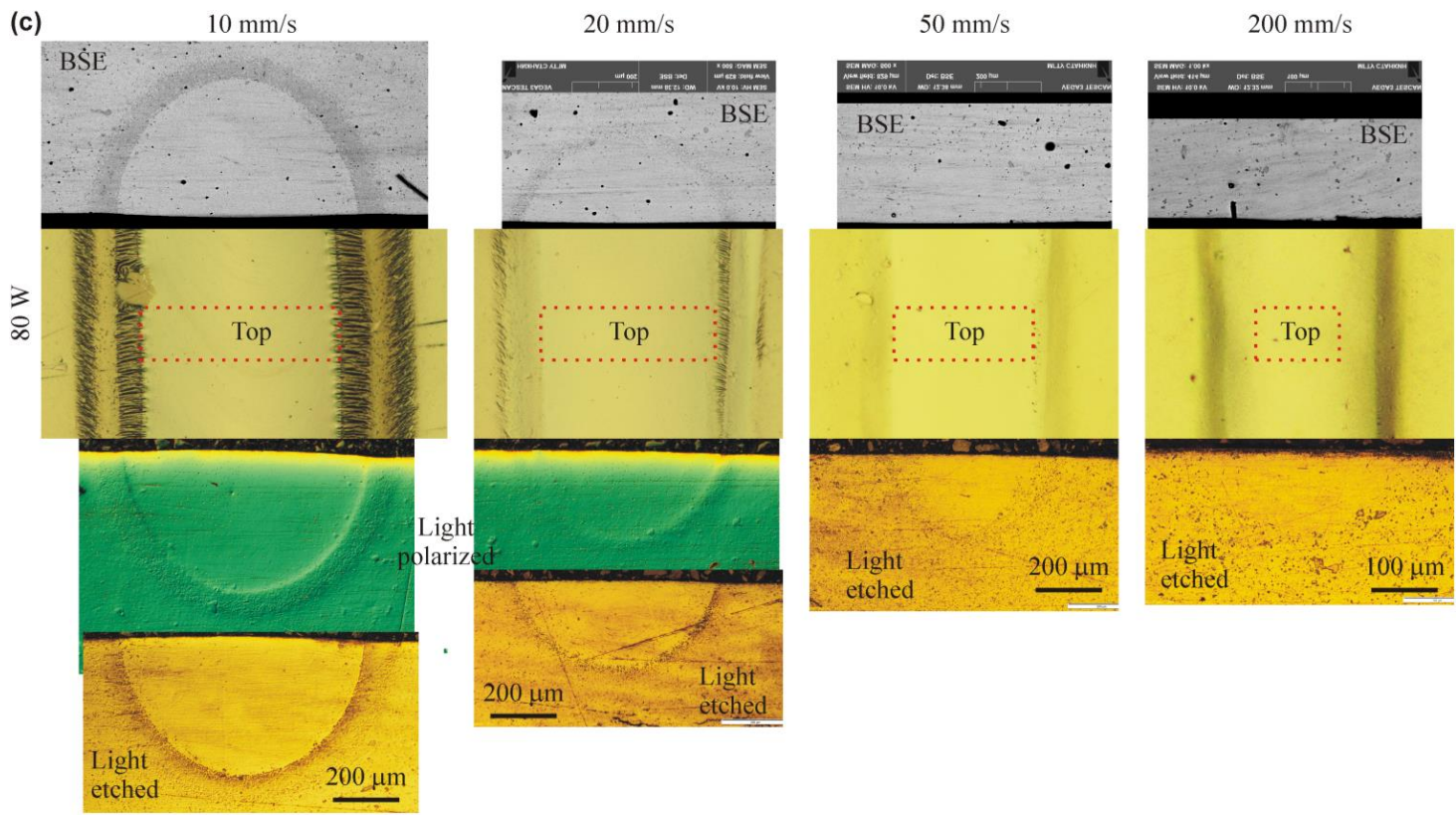

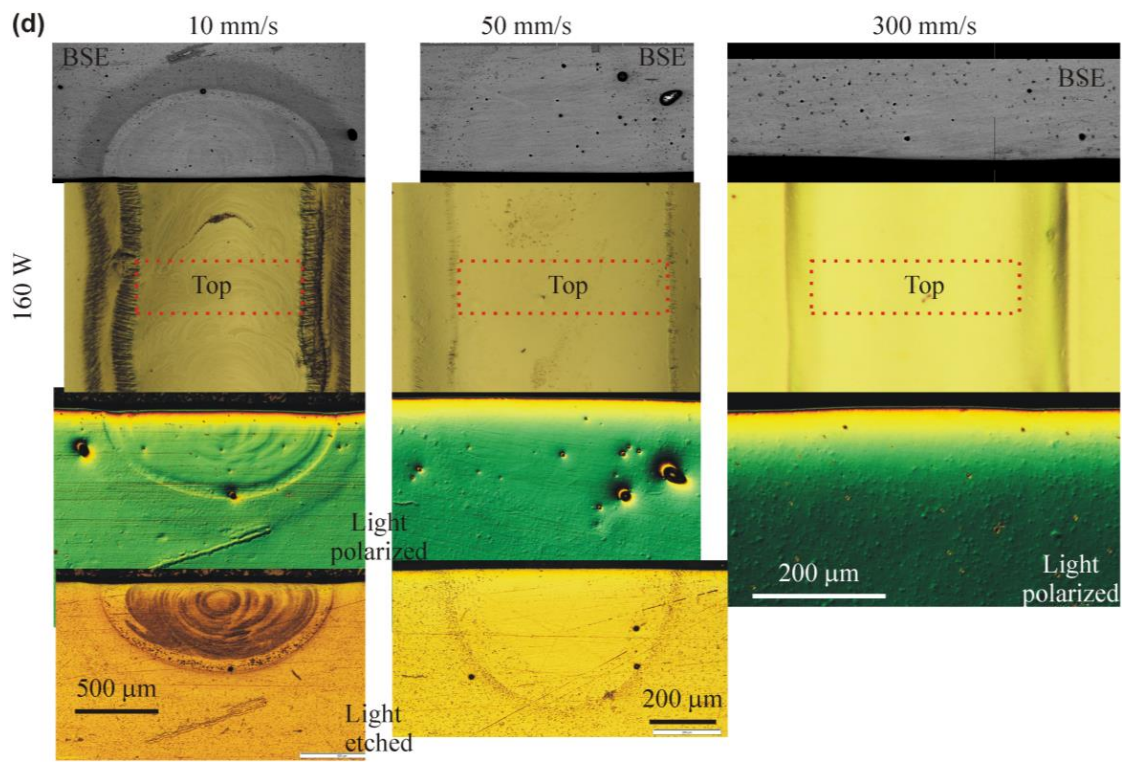

**Fig. 10.** (Continued.)

The crystals can precipitate from the overcooled liquid phase in the temperature interval between the crystallization onset $T_x$ and liquidus $T_l$ points. The issue is whether the crystals grow from the initial amorphous phase in the heat-affected zone of the laser beam where temperature does not overcome the liquidus point or the crystals grow at quenching from the liquid phase formed in the remelted domain where the maximum temperature is above the liquidus point. The former hypothesis seems more likely because the initial amorphous phase contains crystalline inclusions facilitating crystallization at the elevated cooling rates attained in the considered laser processes. In Fig. 10b, the top view at $V = 10$ mm/s shows intersection of the laser-processing zone with a scratch. The scratch disappears in the central bright zone but is clearly visible in the surrounding precipitation zone. This proves that the precipitation zone has not been in liquid state while the bright zone has been completely remelted. Thus, the boundary between the bright and precipitation zones is the remelting boundary where the maximum temperature attains the liquidus point.

The inner zone of columnar crystals is not visible in cross sections. See, for example, SEM and light-microscopy images at $V = 10$ mm/s in Fig. 10c. The SEM images in backscattered electrons indicate that there is no compositional contrast in the inner zone of columnar crystals. This means a diffusionless crystallization mechanism. On the contrary, the outer row of columnar crystals in the top views corresponds to contrast zones revealed by both SEM and light microscopy in cross sections. The columnar crystals become shorter and completely disappear with increasing the scanning speed. See, for examples Fig. 10a at 5 and 10 mm/s and Fig. 10c at 10 and 20 mm/s. The abovementioned figures show that the gap between the two rows of columnar crystals increases with the scanning speed. Thus, at higher scanning speeds, the darker contrast zones around the brighter remelting zone can be due to approximately eqiuaxed precipitates.

In Fig. 10d at $P = 160$ W and $V = 10$ mm/s, multiple concentric vortices are observed in the cross sections and on the top view. Similar vortices were reported in the remelted domains of soda-lime glass after laser processing [26]. They can be traces of convection in the melt pool. At lower power $P$ or at greater scanning speed $V$, the vortices are not revealed. This may indicate that convection in the melt pool becomes less important with decreasing $P$ and increasing $V$. The width of the remelted zone $W$ is measured on the top views as indicated by the dotted-line squares in Fig. 10. Taking into account the observed perfect semicircular shapes in the cross sections, the remelted depth is accepted to be half the width, $H = W/2$. Table 6 summarizes the measurement results.

When the remelted zone is bounded by columnar crystals, the contrast is clear and the uncertainty in $H$ is of the order $\pm 2$ μm. At higher $V$, when the remelted zone is bounded by equiaxed precipitates, the contrast is vague and the uncertainty can increase up to $\pm 5$ μm.

**Table 6.** Measured melt pool depth, $H$ (μm), versus scanning speed $V$ and laser power $P$

| $V$ (mm/s) | $P = 20$ W | $P = 40$ W | $P = 80$ W | $P = 160$ W |
|---|---|---|---|---|
| 5 | 115 | - | - | - |
| 10 | 108 | 192 | 300 | 500 |
| 20 | 98 | 178 | 263 | - |
| 50 | 80 | 140 | 210 | 315 |
| 100 | 74 | 103 | 170 | 236 |
| 200 | 63 | 82 | 126 | 185 |
| 300 | - | - | 110 | 157 |
| 500 | - | - | 90 | 125 |
| 700 | - | - | 84 | 113 |
| 1000 | - | - | 75 | 105 |
| 1500 | - | - | - | 87 |
| 2000 | - | - | - | 76 |
| 3000 | - | - | - | 63 |

## 5. Discussion

The observed depth $H$ and width $W \approx 2H$ of the melt pool listed in Table 6 can be compared to the heat transfer model described in Section 3. The principal difficulty is that the model assumes constant values of material properties while they can substantially vary depending on the amorphous/crystalline state and temperature as shown, for example, in Figs. 8 and 9. The melt pool boundary corresponds to the melting isotherm $T_m$ where the melting point is set to the liquidus temperature $T_l$. Conductive heat transfer to the surrounding solid phase controls the position of this isotherm. Therefore, material properties are taken for the crystalline phase at the melting point. Table 7 lists the accepted model parameters to estimate the size of the melt pool.

**Table 7.** Parameters of the heat transfer model

| Parameter | Value | Source |
| --- | --- | --- |
| Ambient temperature, $T_a$ | 300 K | Experimental condition |
| Melting point, $T_m$ | $T_l = 1115$ K | Table 2 |
| Onset crystallization temperature, $T_x$ | 712 K | Table 2, present experiments |
| Thermal diffusivity, $\alpha$ | 7 mm²/s | Table 5 |
| Thermal conductivity, $\lambda$ | 16 W/(m K) | Table 5 |

Figure 11 plots the experimental data of Table 6 as functions of melt depth $H$ versus scanning speed $V$ at the constant values of laser power $P$. The heat transfer model predicts such functions by the dimensionless curve shown in Fig. 3. The universal dimensionless curve is defined by two scaling parameters, $R_b$ and $V_0$. The goal of Fig. 11 is to find the values of these scaling parameters assuring the best theoretical fit to the experimental data. The scaling parameters are not independent. Relation $R_b V_0 = 2\alpha = 14$ mm²/s follows from Eq. (7). Thus, this is essentially a one-parameter fit. The found values of the scaling parameters are shown on the plots for every value of laser power. The corresponding model curves agree with the experimental points in Fig. 11. The discrepancy is, generally, low but appears to be systematic at higher $V$ in Fig. 11d. The reason can be a systematic error in measuring the melt pool size because of low contrast in the corresponding experimental images.

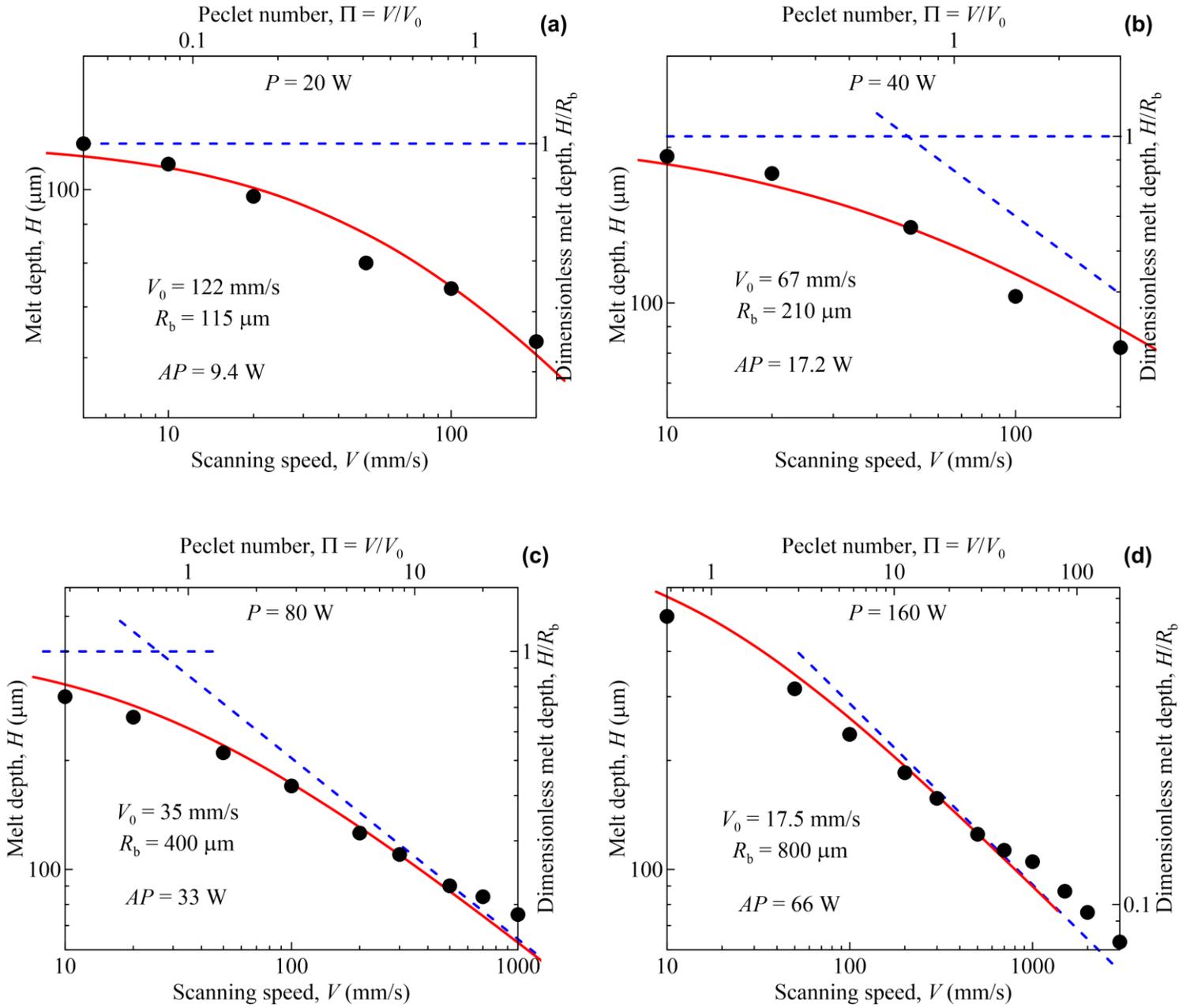

**Fig. 11.** Experimental data (points) fitted by the heat transfer model (full lines) versus scanning speed $V$ at the following values of incident laser power $P$ (W): 20 (a), 40 (b), 80 (c), and 160 (d). Dashed lines are asymptotes of the theoretical curves. Fitting parameters $V_0$ and $R_b$ and the corresponding estimated absorbed power $AP$ are indicated in the plots.

Substituting the estimated values of $R_b$ in Eq. (3) results in the absorbed laser power $AP$. The latter value is indicated in Fig. 11 too. Table 8 lists the effective absorptance $A$ estimated as the ratio of the absorbed power to the incident laser power $P$. The effective absorptance defines the fraction of the incident laser energy transformed into the thermal energy of material and accounts for not only radiation reflection by also evaporation loss. Indeed, the obtained effective absorptance of 0.43 is below the optical absorptivity of 0.57 estimated as the absorptivity of Zr at the room temperature [27].

**Table 8.** Effective absorptance estimated from the theoretical fit to the experimental data

| Laser power, $P$ (W) | 20 | 40 | 80 | 160 | Mean ± MSD[*)] |
|---|---|---|---|---|---|
| Effective absorptance, $A$ | 0.47 | 0.43 | 0.41 | 0.41 | 0.43 ± 0.02 |

[*)] Mean square deviation (MSD)

To estimate the contribution of evaporation to the energy balance at laser processing, one should know that the vapor velocity is limited by its sound speed $c$ [28]. Suppose that overheating at evaporation is not considerable. Then, the vapor pressure is approximately equal to the standard atmosphere $p_s$ and its temperature approaches the boiling point $T_b$ at the atmospheric pressure. In such conditions, the maximum vapor velocity is estimated as the sound speed in the gas of Zr atoms,

$$c = \sqrt{\frac{5}{3}\frac{kT_b}{m}}, \qquad (12)$$

where $k$ is the Boltzmann constant, $m$ the atomic mass, and the factor of 5/3 is the adiabatic exponent of monatomic gas. The maximum evaporation energy loss per unit surface $f$ becomes

$$f = ncL, \text{ with } n = \frac{p_s}{kT_b}, \qquad (13)$$

where $L$ is the latent heat of evaporation per atom and $n$ the vapor number density estimated by the ideal gas law. Table 9 shows the calculation of the evaporation loss with the material properties taken from Ref. [29]. The total evaporation loss depends on the evaporation area. If the evaporation takes place within the projection of the laser beam on the surface, the evaporation area is the circle of 100 μm diameter. The total evaporation loss of 10 W is obtained as the product of the area of this circle multiplied by the specific loss $f$ listed in Table 9. This is a rough order-of-value estimate neglecting, for example, the dependence on the laser power. Nevertheless, this value of total evaporation loss is reasonable and can explain reducing the effective absorptance relative to the optical absorptivity.

**Table 9.** Estimation of the evaporation loss

| Physical quantity | Value | Source |
|---|---|---|
| Boiling point, $T_b$ | 4682 K | [29] |
| Atomic mass, $m$ | 91.22 a.m.u. | [29] |
| Latent heat of evaporation, $L$ | 581.6 kJ/mol | [29] |
| Sound speed, $c$ | 843 m/s | Eq. (12) |
| Number density, $n$ | $1.57 \cdot 10^{24}$ m$^{-3}$ | Eq. (13) |
| Evaporation loss per unit surface, $f$ | 1.28 GW/m$^2$ | Eq. (13) |

The applied experimental methods reveal the width and depth of the melt pool while they cannot measure its length. However, the obtained values of scaling parameters $R_b$ and $V_0$ make it possible a theoretical restoring of the longitudinal melt pool profile. Equation (8) gives a dimensional generatrix profile depending on the Peclet number $\Pi$. Figure 12 shows the examples of the top view on the melt pool. They are theoretically estimated by Eq. (8) using the experimentally obtained scaling parameters. In this equation, the dimensionless temperature $\psi$ is set to 1, that corresponds to the melting point. The melt pool aspect ratio, the ratio of the length to the width, is determined by the Peclet number $\Pi = V/V_0$ while the size is proportional to parameter $R_b$.

Figure 12 indicates that the melt pool shape varies from approximately equiaxed at $\Pi$ of the order of 1 and less to well elongated at $\Pi$ of the order of 100 and greater. In the present experiments, the former conditions correspond to the scanning speed of the order of centimeters per second while the latter conditions are typical at the scanning speed of the order of meters per second. The melt pool width considerably reduces with increasing the scanning speed $V$ at a constant laser power. The distance from the melt pool forward boundary to the laser spot reduces as well while the distance from the backward boundary to the laser spot is equal to the constant value of $R_b$. Thus, the width decreases and the length tends to a constant. The consequence is the considerable elongation of the melt pool with increasing *V*. Simultaneously, the melt pool shifts backward relative to the laser spot.

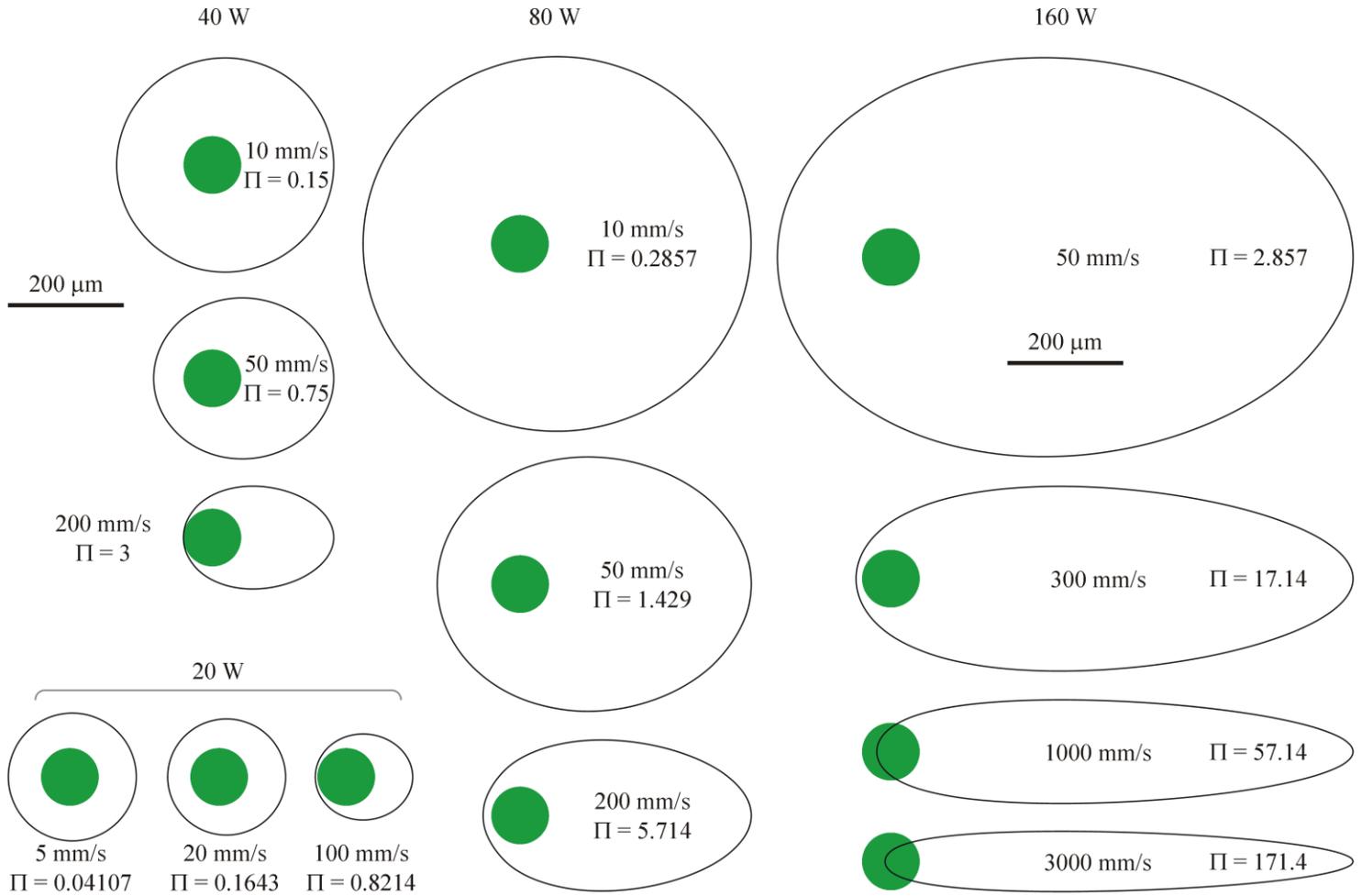

**Fig. 12.** Theoretically estimated top view on the melt pool at laser power *P* (W): 160 (right column); 80 (middle column); 40 (left column); 20 (row at the bottom left). The solid line is the melting isotherm. The green circle is the laser spot. The values of the scanning speed *V* and the corresponding Peclet number Π are indicated near every image.

Considerable evolution of microstructure at laser processing is expected in the temperature interval from the onset crystallization $T_x$ to melting $T_m$ points. The $T_x$ isotherm is given by the same Eq. (8), where the dimensionless temperature is set to

$$\psi = \frac{T_x - T_a}{T_m - T_a}. \qquad (14)$$

Figure 13 superposes the theoretically estimated melting and onset crystallization isotherms on the experimental top views. The melting isotherm (solid line) generally inscribes into the clear band in the middle indicating that this band is a remelted domain. The domain encircled by the melting isotherm is slightly narrower than the remelted band in Fig. 13a while it is slightly wider in Fig. 13c. It seems that such variations are not meaningful because they reflect the difference between the experimentally measured and theoretically predicted remelted width. Indeed, Fig. 13a corresponds to the left point in Fig. 11a, where the thermal model underestimates the observed remelted width. On the contrary, Fig. 13c corresponds to the left point in Fig. 11c, where the thermal model overestimates the remelted width. The $T_x$ isotherm (dashed line in Fig. 13) generally circumscribes the darker zones of precipitation. The domain encircled by this isotherm is systematically wider than the observed precipitation zone. The difference can be explained by the non-equilibrium conditions of laser processing where crystallization may start at a temperature greater than the onset crystallization temperature $T_x$ determined in quasi-equilibrium conditions. Another possible reason is a slow crystal growth rate at temperatures around $T_x$, where crystals are not yet visible on the top-view images.

    Figures 13a-d reveal the bands of columnar crystals. The inner bands a close to the melting isotherms while the outer bands are close to the $T_x$ isotherms. Columnar crystals commonly grow in the direction of temperature gradient. The gradient direction is perpendicular to the isotherm. The growth zone of inner columnar crystals is estimated as zone A in Figs. 13a-d, where the crystals are approximately perpendicular to the melting isotherm. The growth zone of outer columnar crystals is estimated as zone B, where the crystals are approximately perpendicular to the $T_x$ isotherm. Zone B is considerably shifted to the backward side of the melt pool. This indicates a delay in crystal grows in the outer bands. One can explain such a delay by a considerable overcooling reducing the crystal growth rate. Comparison of Figs. 13c-e shows that the columnar crystals become shorter with increasing the scanning speed $V$ and finally disappear. This can be due to shortening the thermal cycle of laser processing with increasing $V$.

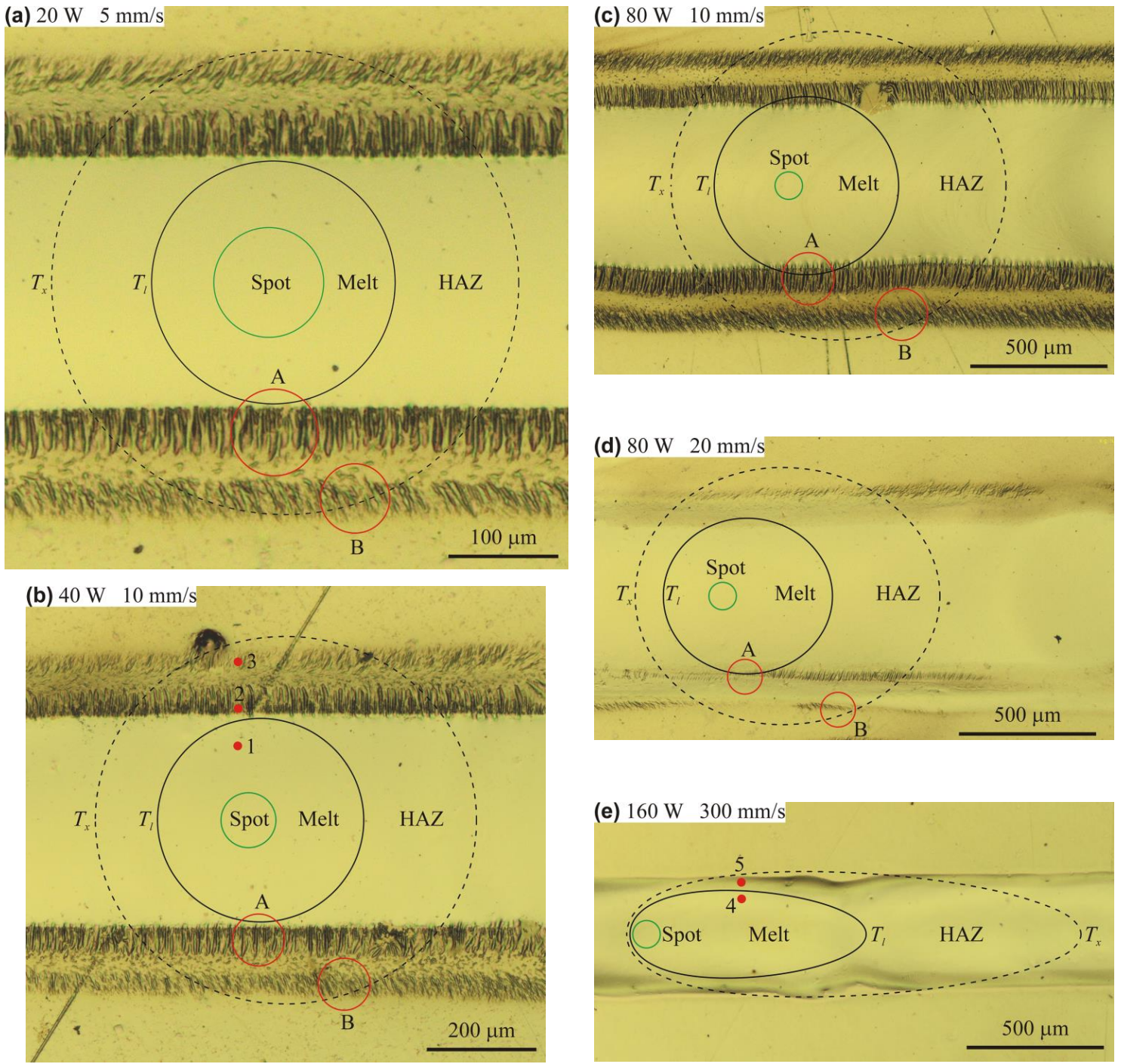

**Fig. 13.** Theoretically estimated melting (black solid line) and onset crystallization (dashed line) isotherms on the experimentally observed top views of the laser-processed surface at the following process parameters: $P = 20$ W and $V = 5$ mm/s (a); $P = 40$ W and $V = 10$ mm/s (b); $P = 80$ W and $V = 10$ mm/s (c); $P = 80$ W and $V = 20$ mm/s (d); $P = 160$ W and $V = 300$ mm/s (e). The green circle designates the laser spot. Red circles A and B encircle the expected domains of columnar crystals formation in the inner and outer rows, respectively. Points 1 – 5 are chosen in the characteristic microstructure domains. Laser scans from right to left.

To understand formation of the observed microstructure patterns, thermal cycles are theoretically calculated in points 1-3 in Fig. 13b and 4-5 in Fig. 13e. The points on the surface have coordinate $Z = 0$ (see Fig. 2). Table 10 shows the chosen values of coordinate $Y$. In the moving frame shown in Fig. 2, coordinate $X$ varies as

$$X = Vt, \quad (15)$$

where $t$ is time. The function of temperature versus time is obtained by substituting the above mentioned values of coordinates X, Y, and Z into Eq. (2). Figure 14 shows the resulting thermal cycles. Parameters $\lambda$ and $\alpha$ are listed in Table 7. Absorbed laser power $AP$ is shown in Fig. 11. Figure 14 shows the resulting thermal cycles.

**Table 10.** Theoretically estimated parameters of thermal cycles.

| Point | Process parameters | $Y/\mu m$ | Microstructure | Crystallization time (ms) | Mean cooling rate (K/s) |
|---|---|---|---|---|---|
| 1 | $P = 40$ W | 135 | Remelted | 23.5 | $1.7 \cdot 10^4$ |
| 2 | $V = 10$ mm/s | 200 | Inner columnar crystals | 55.7 | - |
| 3 |  | 285 | Outer columnar crystals | 36.6 | - |
| 4 | $P = 160$ W | 130 | Remelted | 2.7 | $1.5 \cdot 10^5$ |
| 5 | $V = 300$ mm/s | 190 | Precipitation | 3.2 | - |

Point 1 in Fig. 13b corresponds to curve 1 in Fig. 14. Point 1 is in the mono-phase remelted zone. It is supposed that primary crystalline inclusions completely dissolve when the temperature becomes above the liquidus point $T_l$ on the ascending branch of curve 1 while no crystals are formed in the interval between the onset cryctallization $T_x$ and liquidus temperatures on the descending branch of this curve. The temperature decreases from $T_l$ down to $T_x$ during the time interval of approximately 23.5 ms as shown in Fig. 14. The temperature difference $T_l - T_x$ divided by the time interval gives the mean cooling rate around $1.7 \cdot 10^4$ K/s. This value is much greater than the critical cooling rate of 10 K/s reported for alloy Vit 106 [24]. Therefore, a completely amorphous structure is expected in the remelted zone. The similar processes should occur in point 4 where the thermal cycle is considerably shorter because of much greater scanning speed. Table 10 summarizes the estimated values of crystallization time and mean cooling rate in the crystallization interval. The mean cooling rate is much greater than the critical cooling rate in all the studied regimes.

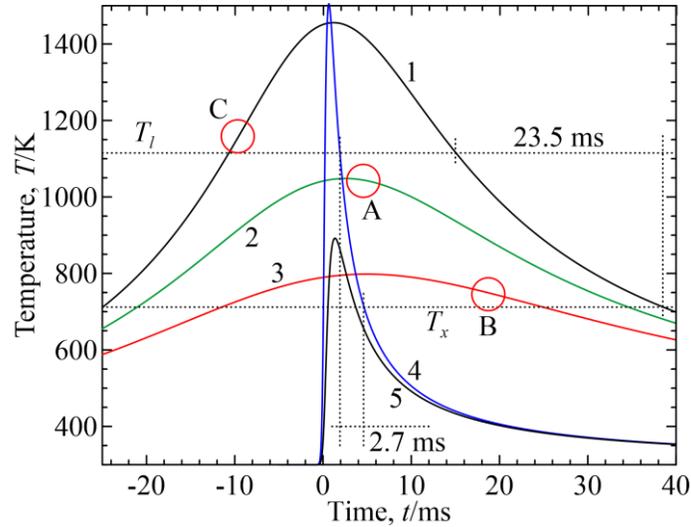

**Fig. 14.** Theoretically estimated thermal cycles in points 1-5: Growth of inner columnar crystals (A); Growth of outer columnar crystals (B); Dissolution of primary crystals (C). Liquidus $T_l$ and onset crystallization $T_x$ temperatures.

Crystalline structure is observed in points 2 and 3 in Fig. 13b. The corresponding temperature curves do not attain the liquidus point. Temperature grows, attains a maximum and decreases. The local cooling rate changes from zero at the maximum temperature to the maximum cooling rate near temperature $T_x$. Therefore, it is not useful to talk about cooling rate. The relevant parameter is the total time interval in the temperature interval from $T_x$ to $T_l$, referred to as the crystallization time. The crystallization times in points 1 – 3 are comparable (see Table 10) while the microstructures are completely different. It is expected that primary crystalline inclusions do not dissolve in points 2 and 3 because material does not melt. These inclusions are nuclei for crystalline phase growth. Thus, one can conclude that the estimated crystallization times of the order of several dozen milliseconds are insufficient for a detectable homogeneous nucleation but sufficient for the observed heterogeneous growth of columnar crystals. The total volume of the heat-affected zone observed in Fig. 13b is of the order of 1 mm³. Since no signs of homogeneous nucleation are found, the mean homogeneous nucleation rate should be of the order or lower than 1/(1 mm³·10ms)

= $10^{12}$ m$^{-3}$s$^{-1}$. The length of columnar crystals in Fig. 13b is around 50 µm. The crystal length divided by the crystallization time is the mean growth rate. Table 11 summarizes the estimated mean values obtained with the crystallization time from Table 10.

**Table 11.** Estimated rates of homogeneous nucleation and crystal growth

|  | Present work, mean value | | | Ref. [30], peak value |
| --- | --- | --- | --- | --- |
|  | Point 1 | Point 2 | Point 3 |  |
| Homogeneous nucleation rate (m$^{-3}$s$^{-1}$) | ≤$10^{12}$ | - | - | 1÷5·$10^{13}$ |
| Crystal growth rate (mm/s) | - | 0.9 | 1.4 | 5 |

The rates of homogeneous nucleation and crystal growth considerably vary with temperature. They are very low at $T_x$ and lower, increase with temperature, attain their maxima, and, finally, decrease down to zero at $T_l$. Table 11 compares the mean values obtained in the present work with the peak values theoretically estimated for a similar Zr-based BMG alloy AMZ4 [30]. It is reasonable that the peak value is several times greater than the mean value. Thus, the obtained data on crystallization kinetics are compatible with the data of Ref. [29]. Crystallization zones A and B in Fig. 13 correspond to the parts A and B of the thermal cycles in Fig. 14. In point 2, the temperature maximum approaches the liquidus point (see Fig. 14). It is expected that crystal growth rate attains its maximum in this thermal cycle. Therefore, the amorphous phase has been consumed when the temperature is near its maximum where crystallization stops. The maximum of thermal cycle 2 in Fig. 14 corresponds to crystallization zone A in Fig. 13b. The inner row of columnar crystals is formed in this zone. In point 3, the temperature maximum is considerably lower (see Fig. 14). It is expected that crystal growth rate does not attain its maximum in this thermal cycle. Therefore, crystallization continues when temperature decreases and approaches $T_x$ (see part B of thermal cycle 3 in Fig. 14). This corresponds to crystallization zone B in Fig. 13b. The outer row of columnar crystals is formed in this zone.

In Fig. 14, thermal cycles 4 and 5 are considerably shorter. Supposing the same mean crystal growth rate of the order of one millimeter per second, one estimates the crystal size in point 5 by multiplying the growth rate by the corresponding crystallization time (see Table 10) resulting in a value around three microns. This is much smaller than the size of primary crystalline inclusions before laser processing visible in cross sections of Fig. 10d. Therefore, crystal growth at $V = 300$ mm/s is not observed in the cross sections but appears as darker precipitation zones on the top views of Figs. 10d and 13e. It is supposed that there are multiple micron-sized or smaller nuclei in the primary material, which are not observed before laser processing. These nuclei can considerably grow in the precipitation zones corresponding to point 5 in Fig. 13e giving rise to a darker contrast on the top view.

The present work has indicated that the cooling rate attained in typical SLM regimes is much greater than the critical cooling rate for the studied Zr-based BMG alloy. It seems that homogeneous nucleation in the amorphous phase is suppressed in these conditions. Nevertheless, laser processing can result in considerable microstructure modification related to crystal growth even at such high cooling rates. The microstructure formation at the laser processing is sensitive to primary crystalline inclusions and nuclei existing in the amorphous phase. The role of these inclusions and their formation in multi-layer additive SLM processes requires further investigation.

## 6. Conclusion

Microstructure formation in $Zr_{57}Cu_{15.4}Ni_{12.6}Al_{10}Nb_5$ Zr-based BMG alloy (trade mark Vit 106) at laser processing is studied for a wide variety of processing regimes typical for SLM additive manufacturing. Single-track experiments are accomplished concerning laser scanning over an amorphous-state substrate at constant laser power $P$ varying from 20 to 160 W and scanning speed $V$ varying from 5 to 3000 mm/s. The results are illustrated using a model of heat transfer in the laser impact zone. Original laser-flash and dilatometry analysis provides with the thermophysical and kinetic parameters necessary for the model.

An essentially axially-symmetric microstructure of laser tracks is observed where the axis is the line drawing by the laser beam axis on the substrate surface. The axial symmetry agrees with the heat transfer model. In the cross section perpendicular to the axis, a central half-circular bright domain is identified as the remelted zone. An annular darker domain encircles the remelted zone. This domain is identified as the crystallization zone.

The model fits the experimentally obtained dependencies of the remelted zone size versus processing parameters $P$ and $V$. The comparison between the model and the experiments indicates that 43 ± 2% of the incident laser energy is transferred into the substrate thermal energy in the studied conditions. The combined energy losses by radiation reflection and material evaporation explain the indicated value of the effective absorption.

The thermal model reconstructs the three-dimensional shape of the melt pool and heat-affected zones as the liquidus $T_l$ and onset crystallization $T_x$ isotherms, respectively. The liquidus isotherm generally inscribes into the remelted zone while the $T_x$ isotherm circumscribes the crystallization zone. The melt pool is approximately equiaxed at the scanning speed of several centimeters per second. It elongates considerably when the scanning speed increases up to several meters per second. The melt pool aspect ratio is related to the Peclet number.

One can distinguish the inner and outer rows of columnar crystals separated by a band of equiaxed inclusions in the crystallization zone at the scanning speed of a few centimeters per second and lower. The inner columnar crystals are perpendicular to the scanning axis while the direction of the outer columnar crystals is oblique. At higher scanning speed, the crystallization zones look like darker bands without internal structure.

The observed microstructure is explained by the thermal cycles predicted by the model. It seems that primary crystalline inclusions existing in the substrate before laser processing dissolve at the laser melting. The mean cooling rate in the remelted zone is much greater than the critical cooling rate. Therefore, homogeneous nucleation is not expected. Nevertheless, the theoretically estimated crystallization times are sufficient for a considerable crystal growth in the heat-affected zone where primary crystalline inclusions and nuclei are not completely dissolved.

**Authors' contribution**   Conceptualization AVG, SNG. Investigation ADK, AYuK, PAP, RSK, TVT. Supervision AVG. Data curation ADK, AYuK. Funding acquisition AVG. Writing—original draft AVG, RSK. Writing—review and editing OBK, SNG, TVT.

**Funding**   This work was supported by the Russian Science Foundation (Grant Agreement No. 21-19-00295, https://rscf.ru/project/21-19-00295/). The experiments were carried out using the equipment of the Centre of collective use of MSUT "STANKIN" supported by the Ministry of Science and Higher Education of the Russian Federation (Contract No. 075-15-2021-695).

**Declarations**

**Ethical approval**     Not applicable.

**Consent to participate**     Not applicable.

**Consent to publication**     Not applicable.

**Conflict of interest**   The authors declare no competing interests.